\documentclass[12pt]{article}
\usepackage{amssymb}
\usepackage{amsmath}
\usepackage{hyperref}
\begin{document}
\title{\bf Transverse Shear Viscosity to Entropy Density for the General Anisotropic  Black Brane in Horava-Lifshitz Gravity}
\author{Mehdi Sadeghi\thanks{ mehdi.sadeghi@abru.ac.ir}  \hspace{2mm} \\
          {\small {\em  Department of Physics, School of Sciences,}}\\
 {\small {\em  Ayatollah Boroujerdi University, Boroujerd, Iran}}\\
    }   
\date{\today}
\maketitle

\abstract{In this paper we calculate the ratio of transverse shear viscosity to entropy density  for the general anisotropic  black brane in Horava-Lifshitz gravity. There is a well-known conjecture that states this ratio should be larger than $\frac{1}{4\pi}$. The ratio of shear viscosity to entropy density is proportional to the inverse square coupling  of quantum thermal field theory,$\frac{\eta }{s} \sim \frac{1}{\lambda^2 }$. Especially in QFT with gravity dual the stronger coupling means the shear viscosity per entropy density gets closer to the lower bound of $\frac{1}{4\pi}$. The KSS bound preserves in the anisotropic scaling model.\\

%\pacs{11.10.Jj, 11.10.Wx, 11.15.Pg, 11.25.Tq}


%--------------------------------------------------------------------------
\section{Introduction} \label{intro}

\indent The AdS/CFT duality  \cite{Ref1,Ref2,Ref3} states
that quantum gravity theories, such as string theory, on an $ AdS_{d+1} $ background is dual to  a $d$-dimensional CFT, which is a non-gravitational theory. Hydrodynamics \cite{Ref4} is an effective theory of QFT at large distances and time-scales.  AdS/CFT duality leads to fluid/gravity duality in this limit\cite{Ref5,Ref7}.The hydrodynamics equations are laws of conservation of energy and
momentum \cite{Ref4,Ref8},
\begin{align}
\nabla _{\mu } T^{\mu \nu } &=0, \\
T^{\mu \nu } &=(\rho +p)u^{\mu } u^{\nu } +pg^{\mu \nu }.
\end{align}

 According to dictionary of AdS/CFT, Black Brane within the bulk is dual to a fluid on the boundary . \\

\begin{table}[h]
 \caption{\footnotesize Relevant equations and parameters in the bulk and on
 the boundary}
 \centering
 \small
\begin{center}
  \begin{tabular} { | c | c |}
    \hline
    Bulk  &  Boundary \\
    \hline
    $E_{MN}=R_{MN}-\frac{1}{2}Rg_{MN}+\Lambda g_{MN}=0 $   & $\nabla _{\mu } T^{\mu \nu }=0$ \\
    \hline
        $ds^{2} =-2u_{\mu}dx^{\mu} dr+r^2[\eta_{\mu \nu}+(\frac{r_+}{r})^4 u_{\mu}u_{\nu}]dx^{\mu}dx^{\nu} $   & $T^{\mu \nu }=(\pi T)^{4}(g^{\mu \nu}+4 u^{\mu}u^{\nu})$ \\
    \hline
  \end{tabular}
   \label{table-tension}
\end{center}
\label{tbl:tablelabel}
\end{table} 

\indent where $ r_+ $, $\Lambda $, $T$ and $u_{\mu }$ are event horizon, cosmological constant, temperature and fluid velocity, respectively \cite{Ref7}.\\

\indent Since the hydrodynamics regime is valid and $\varepsilon =\frac{l_{mfp} }{L}<<1 $, we expand it to the leading order of $\varepsilon $ \cite{Ref5}.

\begin{align}
& T^{\mu \nu } =(\rho +p)u^{\mu } u^{\nu } +pg^{\mu \nu } -\sigma ^{\mu \nu },\\
&\sigma ^{\mu \nu } = {P^{\mu \alpha } P^{\nu \beta } } [\eta(\nabla _{\alpha } u_{\beta } +\nabla _{\beta } u_{\alpha })+ (\zeta-\frac{2}{3}\eta) g_{\alpha \beta } \nabla .u],\\& P^{\mu \nu }=g^{\mu \nu}+u^{\mu}u^{\nu}, \nonumber
\end{align}
\indent where $\eta$, $\zeta $, $\sigma ^{\mu \nu }$ and $P^{\mu \nu }$ are shear viscosity, bulk viscosity, shear tensor and projection operator, respectively \cite{Ref7,Ref9,Ref10,Ref12}. In this work we are interested in calculating transverse shear viscosity by Green-Kubo formula \cite{Ref12,Ref13}. \\
\begin{equation}
\eta =\mathop{\lim }\limits_{\omega \to 0} \frac{1}{2\omega } \int dt\,  d\vec{x}\, e^{\imath\omega t} \left\langle [T_{y}^{x} (x),T_{y}^{x} (0)]\right\rangle =-\mathop{\lim }\limits_{\omega \, \to \, 0} \frac{1}{\omega } \Im G_{y\, \, y}^{x\, \, x} (\omega ,\vec{0}).
\end{equation}

\indent In the following, we consider the general black brane in the Horava-Lifshitz gravity \cite{Mateos:2011tv,Mateos:2011ix,Jahnke:2014vwa,Rebhan:2011vd,Mamo:2012sy}. Then, we calculate the shear viscosity to the entropy density ratio by Green-Kubo formula. The ratio is found to satisfy the conjectured bound $\frac{1}{4\pi}$ for this gravity.
%--------------------------------------------------------------------------

%--------------------------------------------------------------------------

%--------------------------------------------------------------------------

\section{Anisotropic  Black Brane in Horava-Lifshitz Gravity}
\label{sec3}

\indent  We study the model with anisotropic scaling. Horava- lifshitz and Einstein-Hilbert gravity with a scalar field, the so-called “khronon” \cite{Blas:2010hb}  have this property. It help us to understand tricritical point in condense matter physics named by Griffiths \cite{Griffiths}. 
Since, Horava-Lifshitz gravity  is a well known theory, we don't mention to its details. There are several ways to create this theory: It can be obtained by coupling a massive gauge field to Einstein gravity or as solutions of Einstein-axion-dilaton \cite{Mateos:2011ix,Mateos:2011tv}  theories as follows,
\begin{equation}\label{Action}
I=\int d^{5}  x\sqrt{-g} \Big(R-2\Lambda-\frac{1}{2}(\partial \phi)^2-\frac{1}{2}e^{2\phi}(\partial \chi)^2\Big),
\end{equation}
Where $\chi$ and $\phi$ are axion and dilaton fields respectively.  The axion field  has a constant profile in the radial
direction and depends linearly on z, $\chi=az$  .The metric of non-extreme black brane in the context of HL gravity is,
\begin{equation}\label{metricHL1}
ds^{2} =-r^{2\alpha}\,h(r)dt^{2} +\frac{dr^{2} }{r^2\,f(r)}+\frac{r^2}{l^2} b(r)(dx^2+dy^2)+\frac{r^2}{l^2}k(r)dz^2,
\end{equation}
Functions $h(r)$ and $f(r)$ are the blackening factors. Metric (\ref{metricHL1})  is invariant under anisotropic scaling of space-time coordinates $t \to \lambda^{\alpha}t $ , $ x_i \to \lambda x_i $ , $ r \to \frac{r}{\lambda}$  where ${\alpha}$ is the dynamical critical exponent.\\ 
Since, we are considering non-extreme black holes, both functions have simple root at the horizon $r=r_+$, such that $h(r_+)/f(r_+)$ is finite at the horizon.

For later convenience, we introduce the dimensionless coordinate $u=\frac{r_+^2}{r^2}$,
\begin{equation}\label{metricHL3}
ds^{2} =-\frac{r_+^{2\alpha}}{u^{\alpha} l^{2{\alpha}}}H(u)dt^{2} +\frac{l^2 du^2 }{4u^2F(u)} +\frac{r_+^2}{u l^2}B(u)(dx^2+dy^2)+\frac{r_+^2}{u l^2}K(u)dz^2,
\end{equation}
where $l$ is radius of AdS. The solution is clearly isotropic in the $xy$-directions, but not in the $z$-direction.\\
 The temperature and the Hawking-Bekenstein entropy density read as \cite{Ref14},
\begin{eqnarray}
T&=&\frac{1}{4\pi \sqrt{g_{uu} g_{tt} } } \partial_u g_{tt}|_{u=1 } =\frac{r_+^{\alpha}}{2\pi l^{\alpha+1}}\sqrt{\frac{F}{H}}H'|_{u=1}= \frac{r_+^{\alpha}}{2\pi l^{\alpha+1}}\sqrt{F'H'}|_{u=1}, \\
s&=&\frac{4\pi}{V}\int d^{3}x \sqrt{-g}=4\pi\left(\frac{r_+}{l}\right)^{3}B(u=1)\sqrt{K(u=1)}.
\end{eqnarray}
Now to find the transverse shear viscosity, one may consider tensor perturbation of the background metric as $g_{xy} + h_{xy}$ \cite{Ref14,Ref15,Ref16} where $x, y$ are parallel to the brane. Since the brane has translational invariance, we apply Fourier transformation.
\begin{equation}\label{Fourier}
h_x^y(t,u,\vec{x})=G(u)\int{\frac{d^4k}{(2\pi)^4}(h^0)^{y}_{x}(\omega,\vec{k})\exp({\imath \vec{k}\cdot\vec{x}-\imath\omega t})}.
\end{equation}
 Regarding the Green-Kubo formula it is enough to set $\vec{k}=0 $.  Plug the perturbed metric in the action of HL gravity and keep terms up to $G^2$:
\begin{equation}
I_2=\int d^5x \Big(K_1 G'^2 -K_2 G^2\Big),
\end{equation}
where
\begin{align}
K_1&=\sqrt{-g} g^{uu}\nonumber,\\
K_2&=\frac{l^{2\alpha} \sqrt{-g}}{r_{+}^{2\alpha}u^{-\alpha}H(u)}\omega^{2}. \nonumber
\end{align}
then the EoM is,
\begin{equation}\label{EoM}
(K_1 G')' + K_2 G=0.
\end{equation}
Eq.(\ref{EoM})  shows that the perturbation satisfies the Klein-Gordon equation as follows,
\begin{equation} \label{eom-perturbationHL}
\frac{1}{\sqrt{-g} } \partial _{u} (\sqrt{-g} g^{uu} \partial _{u} h_{xy} (t,u,\vec{x}) )+g^{\mu \nu } \partial_{\mu } \partial_{\nu } h_{xy} (t,u,\vec{x}) =0.
\end{equation}
By substituting Eq.(\ref{Fourier}) in Eq.(\ref{eom-perturbationHL})\\
\begin{equation} \label{eq27}
\frac{1}{\sqrt{-g} } \partial _{u} (\sqrt{-g} g^{uu} \partial _{u} G(u) )-g^{tt} \omega ^{2} G(u) =0,
\end{equation}
using (\ref{metricHL3}) we obtain,
\begin{equation} \label{main-eqHL}
 G''(u) +\frac{1}{2}\left(\frac{F'}{F} +\frac{H'}{H}+\frac{2B'}{B}+\frac{K'}{K}-\frac{\alpha+1}{u} \right)G'(u) +\frac{l^{2\alpha+2} \omega^{2} G(u)}{4r_{+}^{2\alpha}u^{2-\alpha}F(u)H(u)}=0,
\end{equation}
 in which $'$ denotes derivative with respect to $u$. This equation is singular at the horizon $u=1$. Thus, first of all, we study the near horizon behavior by using the Taylor expansion,
\begin{eqnarray}
F(u)&\approx& -(1-u)F'(1),  \nonumber\\
H(u)&\approx& -(1-u)H'(1). \nonumber
\end{eqnarray}
then, 
\begin{eqnarray}
F(u)H(u)&\approx& (1-u)^2F'(1)H'(1)= (1-u)^2 \left(\frac{2\pi l^{\alpha+1} T}{r_+^{\alpha}}\right)^2 .
\end{eqnarray}
Therefore Eq. (\ref{main-eqHL}) transforms to, 
\begin{equation} \label{approx-eq}
 G''(u) -\frac{1}{1-u}G'(u) +\frac{\omega^{2} }{16\pi^2T^2}\frac{1}{(1-u)^2} G(u)=0\, .
\end{equation}
The solution to the above equation is $G(u)=(1-u)^{\beta }$, with
\begin{equation} \label{beta}
\beta =\pm \frac{\imath\varpi }{2} ,\, \, \, \, \varpi \equiv \frac{\omega }{2\pi T}
\end{equation}
In the above equation we choose the minus sign, since we are interested in the incoming waves at the horizon. Coming back to the main equation (\ref{main-eqHL}), we consider the following ansatz with an expansion in terms of $\varpi$,
\begin{equation} \label{solutionHL}
G(u)=\widetilde{F}(u)^{\frac{-\imath\varpi }{2}} (\widetilde{h}_0(u)+\frac{\imath\varpi }{2} \widetilde{h}_1(u)+O(\varpi ^{2} )),
\end{equation}
where $\widetilde{F}(u)\equiv \sqrt{F(u) H(u)}$ and we set $\widetilde{h}_0(u)=1$ to normalize $G(u)$ on the boundary.
Plugging (\ref{solutionHL}) into (\ref{main-eqHL}) to first order of $\varpi$, we obtain,
\begin{equation}
 \widetilde{h}_1''+\frac{1}{2}\left(\frac{\widetilde{F}'}{\widetilde{F}} +\frac{2B'}{B}+\frac{K'}{K}-\frac{\alpha+1}{u} \right) \widetilde{h}'_1 -\frac{\widetilde{F}''}{\widetilde{F}} +\frac{\widetilde{F}'}{2\widetilde{F}}\left(-\frac{2B'}{B}-\frac{K'}{K}+\frac{\alpha+1}{u} \right) =0 \,.
\end{equation}
The solution to the above equation is as follows,
\begin{eqnarray} \label{hprimeHL}
\frac{\widetilde{F} \widetilde{h}'_1- \widetilde{F}'}{B^{-1}K^{\frac{-1}{2}}u^{\frac{\alpha+1}{2}}} = C_1 , \\
\label{hsolution}
\widetilde{h}_1 = \log \frac{\widetilde{F}}{C_2}+C_1\int^u \frac{B^{-1}K^{\frac{-1}{2}}u^{\frac{\alpha+1}{2}}}{\widetilde{F}}du ,
\end{eqnarray}
Where $C_1$ and $C_2$ are integration constants. For our purposes the explicit form of $\widetilde{h}_1$ is not important. It would be enough to find $C_1$ by demanding $\widetilde{h}_1$ to be non-singular at the horizon. So we may investigate the near horizon behavior of the integral in (\ref{hsolution})) as follows,
\begin{eqnarray} \label{aaaHL}
\widetilde{F} &\approx& -(1-u) \widetilde{F}'(1)=-(1-u)\frac{2\pi l^{\alpha+1} T}{r_+^{\alpha}} \\
\label{bbb}
\widetilde{h}_1 &\approx& \log \frac{1-u}{C_2}-\frac{C_1r_+^{\alpha}(B(u=1))^{-1}(K(u=1))^{\frac{-1}{2}}}{2\pi l^{\alpha+1} T}\log (1-u) .
\end{eqnarray}
Considering a non-singular $\widetilde{h}_1$ at the horizon, $C_1$ read as , 
\begin{equation}
C_1= \frac{2\pi l^{\alpha+1}T}{r_+^{\alpha}}B(u=1)\sqrt{K(u=1)}
\end{equation}

Following the prescription given in \cite{Ref17,Ref18,Son}, the retarded Green's function will be:
\begin{eqnarray} \label{GreenHL}
G_{y y} ^{x x} (\omega ,\vec{0})&=&-\sqrt{-g} g^{uu} \, G^{*} (u)\, \partial _{u} G(u)|_{u \to 0} \nonumber\\
 &=& \frac{\imath \, r_+^{\alpha+3} \varpi}{\ell^{\alpha+4}} \left[ \frac{ \widetilde{F}'-\widetilde{F} \widetilde{h}'_1}{B^{-1}K^{\frac{-1}{2}}u^{\frac{\alpha+1}{2}}}\right]\Bigg|_{u \to 0}
  \nonumber\\
 &=& -\frac{\imath\, \omega \, r_{+}^{\alpha+3} }{2\pi \, T l^{\alpha+4} } C_1
\end{eqnarray}
where in the last line Eq.(\ref{hprimeHL}) was used.

\noindent Now the transverse shear viscosity can be found by using Green-Kubo formula \cite{Ref16},
\begin{equation} \label{eq42}
\eta_{\perp}=\eta_{yy}^{xx}=-\mathop{\lim }\limits_{\omega \to 0} \frac{1}{\omega } \Im G_{y y} ^{x x} (\omega ,\vec{0})=\,\left(\frac{r_+^3}{l^3}\right)B(u=1)\sqrt{K(u=1)}
\end{equation}

\noindent Then the ratio of transverse shear viscosity to entropy density is,
\begin{equation} \label{etapers}
\frac{\eta_{\perp}  }{s}=\frac{1}{4\pi } .
\end{equation}
 \noindent The importance of this result is that it is an exact and universal result regardless of details of black-brane solution. There are two kind of shear viscosity in anisotropic black brane solution: transverse and longitudinal. In this paper we calculate transverse shear viscosity.

%--------------------------------------------------------------------------

 \section{Results and discussion}

\noindent We prove the universality of the $\frac{\eta_{\perp}}{s}$ for the general anisotropic black brane solution in Horava-Lifshitz gravity. This value for QGP is the same as experimental data. This value is an example for supporting string theory.
 
%--------------------------------------------------------------------------
\section{Conclusion}

\noindent We showed that the lower bound of the ratio $\frac{\eta_{\perp}}{s}$ preserves for the general anisotropic black brane in Horava-Lifshitz gravity.  This bound is known as KSS conjecture \cite{Ref16} and considered for strongly interacting systems where reliable theoretical estimate of the viscosity is not available. It tells us that the ratio $\eta/s$ has a lower bound, $\frac{\eta }{s} \ge \frac{\hbar }{4\, \pi \, k_{B} } $, for all relativistic quantum field theories at finite temperature without chemical potential and can be interpreted as the Heisenberg uncertainty principle \cite{Ref14,Ref16}. The massive term with the Dirichlet boundary condition and regularity on the horizon {\cite{Hartnoll:2016tri,Sadeghi:2018ylh,Sadeghi:2018vrf} violate the KSS bound but the massive term with the Petrov-like boundary condition preserve this bound \cite{Pan:2016ztm}. However, this conjecture violates for higher derivative gravities 
\cite{Ref21,Ref22,Ref23,Sadeghi:2015vaa, Parvizi:2017boc, Burikham:2016roo, Ge:2009ac,Ge:2015owa, Cai:2008ph, Li:2017ncu, Dehghani:2013ldu,Wang:2016vmm}.
%--------------------------------------------------------------------------

%--------------------------------------------------------------------------
\vspace{1cm}
\noindent {\large {\bf Acknowledgment} }   Author would like to thank Shahrokh Parvizi and the referee of INJP for useful comments and suggestions.

%--------------------------------------------------------------------------

\end{document}